\documentstyle[twoside,fleqn,espcrc2,epsf]{article}

\newcommand{\hs}{\hspace{0.1cm}}

\title{Chiral measurements with the Fixed-Point Dirac operator
  and construction of chiral currents}
\author{P.\ Hasenfratz\address[BERN]{Institute for Theoretical Physics,
    University of Bern, Sidlerstrasse 5, CH-3012 Bern,
    Switzerland}\thanks{Based on a talk by T.J. and a poster by K.H.}, 
    S.\ Hauswirth\addressmark, K.\ Holland\addressmark, 
    T.\ J\"org\addressmark, F.\ Niedermayer\addressmark}

\begin{document}

\begin{abstract}
In this preliminary study, we examine the chiral properties of the
parametrized Fixed-Point Dirac operator $D^{\rm FP}$, see how to improve its
chirality via the Overlap construction, measure the renormalized quark
condensate $\hat{\Sigma}$ and the topological susceptibility $\chi_t$, and
investigate local chirality of near zero modes of the Dirac operator. We also
give a general construction of chiral currents and densities for chiral
lattice actions. 
\end{abstract}

\maketitle

\section{Introduction}
There has been enormous recent interest in lattice QCD with chiral fermions,
with several approaches now being used in simulations \cite{Vra01}. We use one
approach, Fixed-Point actions \cite{Has94}, which have many desirable features,
including chiral symmetry. This preliminary study gives the first chiral
measurements with this action. We also show how to construct chiral densities
and conserved currents for chiral actions.

\section{Chiral properties of $D^{\rm FP}$}
\begin{figure}[th]
\epsfxsize=7cm
\epsfbox{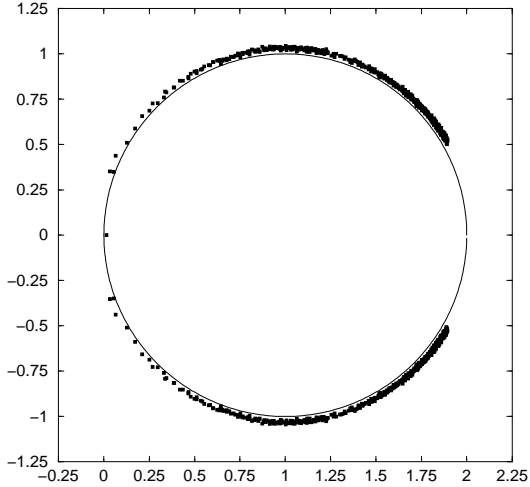}
\caption{Eigenvalue spectrum of $D^{\rm FP}$.}
\label{fig:spectrum}
\end{figure}

A lattice Dirac operator $D$ which satisfies the Ginsparg-Wilson (GW) relation \cite{Gin82}
\begin{equation}
\{D^{-1}, \gamma_5\} = 2 a R \gamma_5,
\end{equation}
has exact chiral symmetry at non-zero lattice spacing $a$ \cite{Lus98}. The
exact Fixed-Point Dirac operator satisfies exactly the GW relation \cite{Has98}
with a non-trivial local function $R$. The function $2R$ can be absorbed into
the definition of $D$ \cite{Has01b} and, for notational simplicity, we shall write $2R=1$ in
most of the following equations. The parametrized Fixed-Point
Dirac operator $D^{\rm FP}$, as described in \cite{Has01}, is an approximate solution
of the GW relation. In Fig.~\ref{fig:spectrum}, we show the eigenvalues of
$D^{\rm FP}$ for Fixed-Point gauge action $S^{\rm FP}_g$ configurations
\cite{Nie01} of volume $4^4$ and $5^4$ with lattice spacing $a \approx 0.16 \hs {\rm fm}$ (as
measured by the Sommer parameter $r_0$). For an exact solution $D$ of the GW
relation, the complex eigenvalues of $D$ lie on the circle of radius 1 and
center at (1,0). As the eigenvalues of $D^{\rm FP}$ lie very close to the GW
circle, we see that $D^{\rm FP}$ satisfies the GW relation well.

\begin{figure}[th]
\epsfxsize=\hsize
\epsfbox{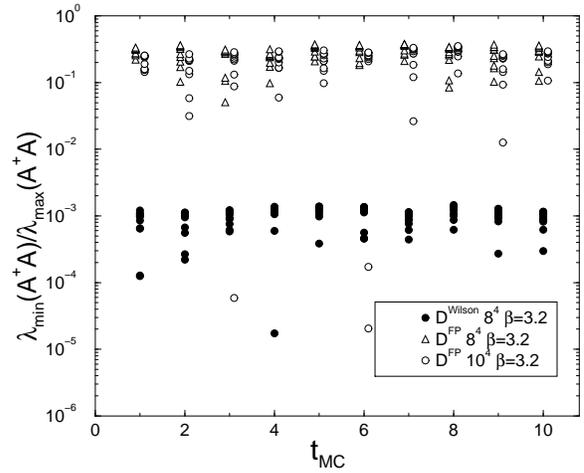}
\caption{Chirality breaking of $D^{\rm Wilson}$ and $D^{\rm FP}$.}
\label{fig:AdaggerA}
\end{figure}

The GW relation is equivalent to
\begin{equation}
A^{\dagger} A = 1, \hspace{0.5cm} A = 1 - D.
\end{equation}
In Fig.~\ref{fig:AdaggerA}, we plot the 10 smallest eigenvalues of
$A^{\dagger} A$ divided by the largest eigenvalue of $A^{\dagger} A$ for 10
Fixed-Point gauge configurations with lattice spacing $a \approx 0.13 \hs {\rm
  fm}$. The largest eigenvalue is $\sim$ 1.5 and 38 for $D^{\rm FP}$ and $D^{\rm
  Wilson}$ respectively. We see that the $A^{\dagger} A$ eigenvalues are much
closer to 1 using $D^{\rm FP}$ than by using $D^{\rm Wilson}$. For larger
volumes, $A^{\dagger} A$ is more likely to have a few very small eigenvalues,
but the vast bulk of the eigenvalues is close to 1.

\begin{figure}[th]
\epsfxsize=7cm
\epsfysize=6cm
\epsfbox{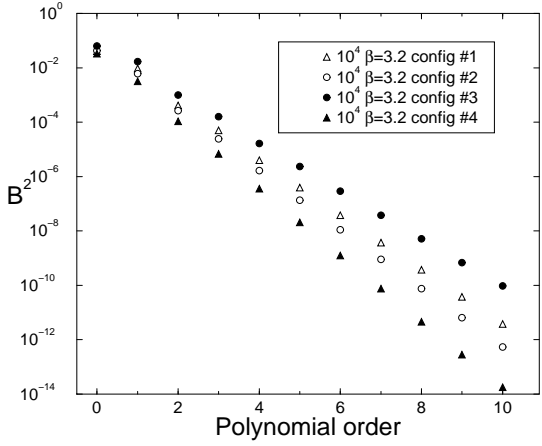}
\caption{Breaking of Ginsparg-Wilson relation.}
\label{fig:gwbreaking}
\end{figure}

\begin{figure}[bth]
\epsfxsize=\hsize
\epsfysize=6cm
\epsfbox{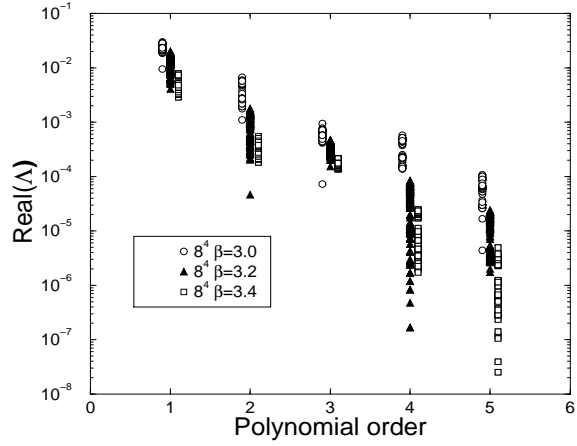}
\caption{Deviation of $D^{\rm Overlap}$ eigenvalues from GW circle.}
\label{fig:stereo}
\end{figure}

An exact solution of the GW relation is given by the Overlap operator \cite{Nar93}
\begin{equation}
D^{\rm Overlap} = 1 - A/\sqrt{A^{\dagger} A}, \hspace{0.5cm} A = 1 - D_0.
\end{equation}
The already very good chiral behavior of $D^{\rm FP}$ can be made exact by
using the Overlap construction. Very precise chiral symmetry is required for
some measurements, such as the chiral condensate. Using $D^{\rm FP}$,
$A^{\dagger} A$ is close to 1 and we use a polynomial approximation of
$1/\sqrt{A^{\dagger} A}$, which should converge rapidly. In
Fig.~\ref{fig:gwbreaking}, we see that the breaking of the GW relation, given by 
\begin{equation}
B^2 = |(D + D^{\dagger} - D^{\dagger} D).v|^2,
\end{equation}
where $v$ is a random vector, falls off exponentially as we
increase the order of the polynomial approximation of $1/\sqrt{A^{\dagger}
  A}$ and becomes $\leq {\cal O}(10^{-10})$ at polynomial order 10 for this
volume. If the eigenvalues $\lambda$ of $D^{\rm Overlap}$ lie exactly on the GW
circle, then $\Lambda=\lambda/(1-\lambda/2)$ is purely imaginary. In
Fig.~\ref{fig:stereo}, we see that ${\rm real}(\Lambda)$ falls off
exponentially as we increase the polynomial order, as the eigenvalues
$\lambda$ get closer and closer to the GW circle. The most rapid
decrease is at the finest lattice spacing $a \approx 0.10 \hs {\rm fm}$.

\begin{figure}[th]
\epsfxsize=\hsize
\epsfbox{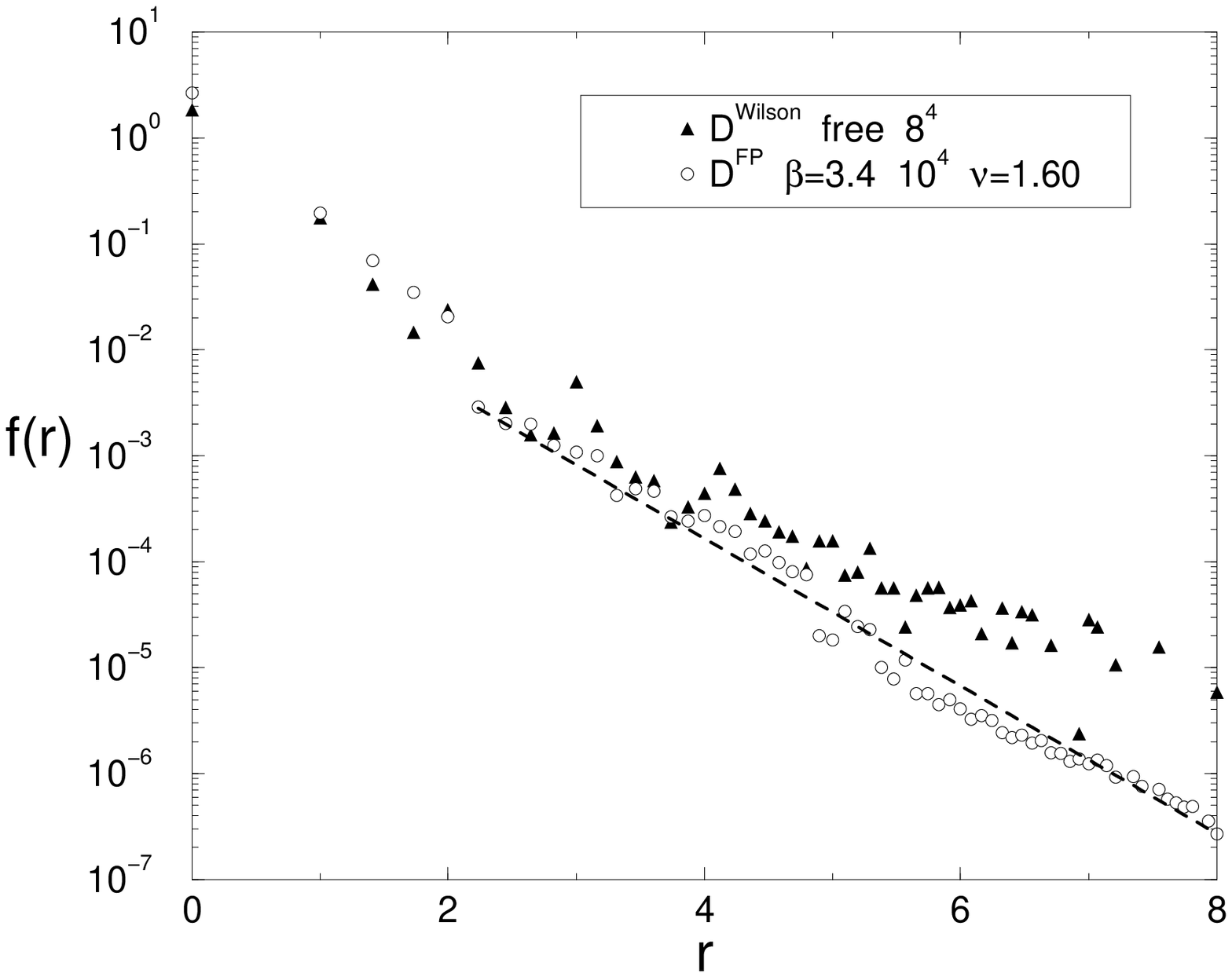}
\caption{Locality of $D^{\rm Overlap}$ using $D^{\rm Wilson}$ and $D^{\rm
    FP}$ (exponential fall-off with $\nu \sim 1.6$).} 
\label{fig:locality}
\end{figure}

A lattice action must be local for the continuum results to be universal
i.e.\hs independent of the details of the lattice action. Locality means that
fields at large separation have an exponentially small coupling. Optimizing
the locality is essential so that e.g.\hs the exponential fall-off of correlation
functions can be separated from direct couplings in the lattice action. The
locality of a lattice Dirac operator can be measured by
\begin{equation}
f(r) = \stackrel{y}{\rm max}\{ |D.v|, ||y-x||=r \},
\end{equation}
where $v$ is a vector with point source at $x$ and $r$ is the square norm. In
Fig.~\ref{fig:locality}, we see that the Overlap operator constructed from
the interacting $D^{\rm FP}$ is more local than by using the free $D^{\rm
  Wilson}$, which in turn is more local than by using the interacting $D^{\rm
  Wilson}$. 

\section{Chiral condensate}
For a theory of $N_f \ge 2$ massless quark flavors, the chiral symmetry is
spontaneously broken if the chiral condensate $\langle \bar{\psi} \psi
\rangle \ne 0$. Using chiral perturbation theory or random matrix theory, the
chiral condensate at finite quark mass and volume has been calculated in the
continuum, both for quenched and full QCD \cite{Osb99}. The quenched QCD condensate
$\Sigma_{m,V,Q} = - \langle \bar{\psi} \psi \rangle_{m,V,Q}$ in topological
sector $Q$ at small $m$ is
\begin{eqnarray}
&& \hspace{-0.75cm} \Sigma_{m,V,Q} =  |Q|/mV + \\
&& \hspace{-0.75cm} m V \Sigma^2 [I_{|Q|}(z) K_{|Q|}(z) + I_{|Q|+1}(z) K_{|Q|-1}(z)]
\nonumber ,
\end{eqnarray}
where $I_Q$ and $K_Q$ are modified Bessel functions, $z=m \Sigma V$ and
$\Sigma=-\langle \bar{\psi} \psi \rangle$ is the quantity we want to
measure. The term $|Q|/mV$ comes from the zero modes, whose contribution is
not suppressed by the fermion determinant in quenched QCD at finite
volume. Subtracting the contribution of the zero modes, for $Q\ne0$
\begin{equation}
\lim_{m \rightarrow 0} \frac{\Sigma^{\rm sub}_{m,V,Q}}{m} = \frac{\Sigma^2
  V}{2 |Q|}.
\end{equation}
By measuring this ratio for several masses, volumes and topological sectors,
the scaling behavior can be used to extract $\Sigma$ \cite{Her99}.

The subtracted condensate is measured by
\begin{eqnarray}
&& \hspace{-0.6cm} \Sigma^{\rm sub}_{m,V,Q} = {\rm Tr}^{\rm sub}[ (D(m) 2R)^{-1} - 1/2 ]
\nonumber \\
&& \hspace{-0.6cm} D(m) 2R = (1-m/2) D(0) 2R + m,
\end{eqnarray}
where $D$ and $R$ are related via the GW relation. We use the Overlap
construction with $D^{\rm FP}$ (hence the Overlap $D^{\rm FP}$) so that the
chiral symmetry is precise enough to go to quark mass $ma=10^{-4}$. The trace
is measured stochastically using random $Z(2)$ vectors. The topology $Q$ is
determined from the chirality of the lowest eigenmodes of $D^{\dagger} D$,
found with an Arnoldi solver. For the topology scan, a Legendre polynomial of
order 2 is used to approximate $1/\sqrt{A^{\dagger} A}$. As a byproduct, we
determine the topological susceptibility from an ensemble of 200 $10^4$
configurations to be $r_0^4 \chi_t = 0.0612(75)$ corresponding to $\chi_t =
(196 \pm 6 \hs {\rm MeV})^4$. In the volumes we examine, all $Q$ zero modes have the
same chirality and their contribution to the condensate is subtracted by using
random $Z(2)$ vectors whose chirality is opposite to those of the zero modes.

\begin{figure}[t]
\epsfxsize=\hsize
\epsfbox{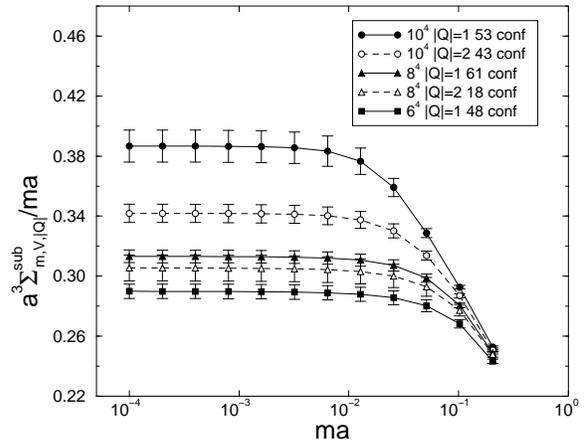}
\caption{$a^3 \Sigma_{m,V,|Q|}/ma$ vs. $ma$ for several volumes and topological sectors.}
\label{fig:sigma_vs_m}
\end{figure}

\begin{figure}[bt]
\epsfxsize=\hsize
\epsfbox{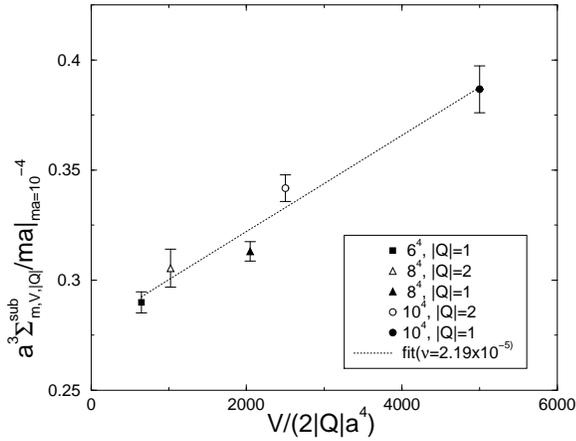}
\caption{$a^3 \Sigma_{m,V,|Q|}/ma$ at $ma=10^{-4}$ vs. $V/(2|Q|a^4)$ and fit to
  finite-size scaling behavior.}
\label{fig:sigma_vs_V}
\end{figure}

We measure the condensate in volumes $6^4,8^4$ and $10^4$ in topological
sectors $|Q|=1,2$ at lattice spacing $a \approx 0.13 \hs {\rm fm}$. The statistics
are given in Table 1. We use 10 random $Z(2)$ vectors to measure the trace for
each configuration and a multi-mass solver to invert $D$ simultaneously at a
number of masses. We use Legendre polynomials of order 5, 7 and 10 to
approximate $1/\sqrt{A^{\dagger} A}$ for volumes $6^4,8^4$ and $10^4$
respectively. This gives sufficiently precise chiral symmetry --- increasing
the polynomial order further, the relative change in $\Sigma^{\rm sub}_{m,V,Q}/m$ is
$\leq {\cal O}(10^{-4})$. The 10 smallest $A^{\dagger} A$ eigenvalues are projected out
and treated exactly. In Fig.~\ref{fig:sigma_vs_m}, we plot $a^3 \Sigma^{\rm
  sub}_{m,V,Q}/ma$ as a function of $ma$ for several volumes and topological
sectors. We see that the data reach a plateau at small quark
masses. Including lattice artifacts, the finite-size scaling behavior is
\begin{equation}
\frac{\Sigma^{\rm sub}_{m,V,Q}}{m} = \frac{\Sigma^2 V}{2
  |Q|} + \frac{c_1}{a^2} + \frac{c_2 m}{a} + ...,
\end{equation}
where $c_1,c_2,...$ are unknown coefficients which have to be fitted. It is
natural to assume that these coefficients, dependent on the UV fluctuations,
are independent of the topological charge $Q$. As shown
in Fig.~\ref{fig:sigma_vs_V}, we fit the data at $ma=10^{-4}$ and determine
that the bare condensate is $a^3 \Sigma = 4.68(26) \times 10^{-3}$. We can
convert this using the Sommer parameter ($r_0/a=3.943(60)$ has previously been
measured at this bare coupling), which gives $r_0^3 \Sigma = 0.287(16)(13)$,
where the first error is statistical and the second the uncertainty in the scale.

\begin{figure}[tb]
\epsfxsize=7cm
\epsfysize=6cm
\epsfbox{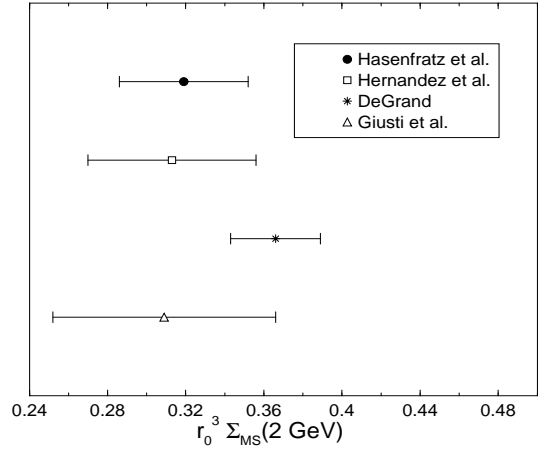}
\caption{Comparison of measurements of $r_0^3 \Sigma_{\overline {\rm MS}}(2
  \hs {\rm GeV})$ by different groups.}
\label{fig:condensate_results}
\end{figure}

The renormalization factor $Z_S$ for the chiral condensate is related to the
quark mass renormalization by $Z_S = 1/Z_m$. The
mass renormalization can be determined from hadron spectroscopy measurements,
as done in \cite{Her01}. Hadron mass measurements using the Overlap $D^{\rm
  FP}$ have been done at the same lattice spacing $a \approx 0.13 \hs {\rm
  fm}$ \cite{Hau01}, giving $Z_S=0.80(6)$ in the renormalization group invariant
scheme. Using this, the renormalized chiral condensate is $r_0^3
\hat{\Sigma} = 0.230(13)(10)(17)$, with statistical, scale and renormalization
errors respectively. This gives $\hat{\Sigma}=(242 \pm 9 \hs {\rm MeV})^3$, or
in the $\overline{\rm MS}$ scheme, $\Sigma_{\overline{\rm MS}}(2 \hs {\rm
  GeV})=(270 \pm 10 \hs {\rm MeV})^3$. Comparing our measurement with other
recent determinations of the renormalized chiral condensate
\cite{Her01,DeG01}, we see in Fig.~\ref{fig:condensate_results} that there is
good agreement.

\begin{table}
\begin{center}
\begin{tabular}{|c|c|c|}
\hline 
$V$ & $|Q|$ & $N_{\rm conf}$ \\
\hline \hline
$6^4$ & 1 & 48 \\ \hline
$8^4$ & 1 & 61 \\ \hline
$8^4$ & 2 & 18 \\ \hline
$10^4$ & 1 & 53 \\ \hline
$10^4$ & 2 & 43 \\ \hline
\end{tabular}
\end{center}
\caption{Statistics for the condensate measurement.}
\end{table}

\section{Local chirality of near zero modes}
From instanton physics, it is expected that near zero modes of a lattice Dirac
operator are localized around instantons and anti-instantons and that, in
these regions, the modes are predominantly either left- or right-handed. These
near zero modes form a finite density $\rho(0)$ as $V \rightarrow \infty$,
breaking the chiral symmetry via the Banks-Casher relation $\langle
\bar{\psi} \psi \rangle = - \pi \rho(0)$. A measure of the local chirality at
lattice site $x$ is \cite{Hor01}
\begin{equation}
\tan[\frac{\pi}{4}(1+X(x))] = \sqrt{\frac{{\psi_{\rm L}}^\dagger \psi_{\rm L}
    (x)}{{\psi_{\rm R}}^\dagger \psi_{\rm R} (x)}}. 
\end{equation}
If the near zero modes are localized and locally chiral, then where
$\psi^\dagger \psi(x)$ is large, $X$ should be close to $\pm 1$. If the modes
are not locally chiral, then $X$ should be close to 0.

\begin{figure}[th]
\epsfxsize=\hsize
\epsfbox{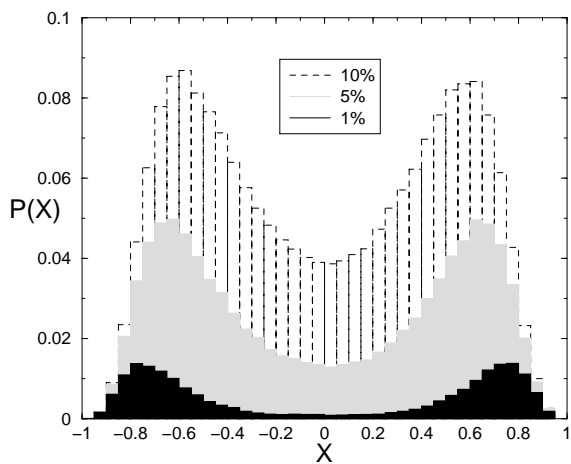}
\caption{Distribution $P(X)$ for top 1\%,5\%,10\% lattice sites $x$ with
  largest $\psi^\dagger \psi(x)$.}
\label{fig:localchirality}
\end{figure}

We analyzed the 10 smallest near zero modes of the Overlap $D^{\rm
FP}$ for 60 $10^4$ configurations at lattice spacing $a \approx 0.13 \hs {\rm
fm}$, using a Legendre polynomial of order 2 to approximate
$1/\sqrt{A^\dagger A}$ (these modes were already used to determine the
topology $Q$ of the gauge configurations). In Fig.~\ref{fig:localchirality},
we plot the distribution $P(X)$, where we include the top 1\%,5\% and 10\%
lattice sites $x$ where $\psi^\dagger \psi(x)$ is largest. We see a clear
double-peaked distribution, with peaks far from $X=0$. Where the modes are
most localized, they are also very chiral. This supports the picture of
instanton dominance of the near zero modes, in agreement with \cite{DeG01b}, unlike
the original claim of \cite{Hor01}.

\section{Construction of chiral currents}
For a chiral lattice action, the advantages of using the conserved currents
are that $Z_V = Z_A = 1$, the current and densities are automatically ${\cal
  O}(a)$ improved and expressions have the same form as in the continuum. The
axial-vector current with the Overlap operator was already given in \cite{Kik99}.
Here we define conserved currents (obtained via the Noether procedure) which
are easy to use in numerical simulations. The vector and axial-vector currents
are
\begin{eqnarray}
V^a_\mu(x) &=& \bar{\psi} \tau^a [P_L K_\mu(x) \hat{P}_R + P_R K_\mu(x) \hat{P}_L] \psi
\nonumber \\
A^a_\mu(x) &=& \bar{\psi} \tau^a [P_L K_\mu(x) \hat{P}_R - P_R K_\mu(x) \hat{P}_L] \psi
\nonumber \\
&& \phantom{.} \hspace{-2cm} P_{R,L} = (1 \pm \gamma_5)/2 \hspace{0.5cm} \hat{P}_{L,R} = P_{L,R} \pm \gamma_5 D/2.
\end{eqnarray}
These currents are conserved and have the correct properties under global vector
and axial-vector transformations e.g.
\begin{eqnarray}
&& \hspace{-0.5cm} i \frac{\delta}{\delta \epsilon^a} V^b_\mu(x) = i f^{abc} A^c_\mu(x)
  \hspace{0.3cm} 
\mbox{axial-vector} \nonumber \\
&& \hspace{-0.5cm} i \frac{\delta}{\delta \epsilon^a} V^b_\mu(x) = i f^{abc} V^c_\mu(x) 
  \hspace{0.3cm} {\rm vector} 
\end{eqnarray}
The kernel $K_\mu(x,U)$ which appears in the currents is defined by
\begin{eqnarray}
K_\mu(x,U) &=& - i \frac{\delta D(\tilde{U}_\mu(x))}{\delta \alpha_\mu(x)}|_{\alpha=0}
\nonumber \\
\tilde{U}_\mu(x) &=& {\rm e}^{i \alpha_\mu(x)} U_\mu(x),
\end{eqnarray}
The kernel can be calculated easily --- the gauge link
$U_\mu(x)$ receives an infinitesimal $U(1)$ rotation and the Dirac operator is
calculated on this new and original gauge configuration (differing only at one
link). The ratio of the difference of these two Dirac operators to the $U(1)$
variation gives $K_\mu(x,U)$. Details of the derivation and properties of the
currents and densities will be given in \cite{Has01b}. 

\section{Summary}
The parametrized Fixed-Point Dirac operator $D^{\rm FP}$ has very good chiral
properties and low-order polynomial approximations can be used in the Overlap
construction to make the chiral symmetry very precise. The Overlap operator is
also much more local using $D^{\rm FP}$ than by using $D^{\rm Wilson}$. Using
finite-size scaling, the bare chiral condensate is measured as $r_0^3 \Sigma =
0.287(16)(13)$ and, estimating the renormalization factor from hadron
spectroscopy, the renormalization group invariant condensate is $r_0^3 \hat{\Sigma} =
0.230(13)(10)(17)$, i.e.\hs $\hat{\Sigma}=(242 \pm 9 \hs {\rm MeV})^3$. In
addition, the topological susceptibility is $\chi_t=(196 \pm 6 \hs {\rm MeV})^4$. Near
zero modes of the Dirac operator are locally chiral, which supports the
picture that they are localized on instantons. We give a general and practical
construction of chiral currents and densities, to be used in e.g.\hs determining
renormalization factors and decay constants.

\section{Acknowledgements}
This work is supported in part by the Schweizerischer Nationalfonds and the
European Community's Human Potential Programme under contract
HPRN-CT-2000-00145, Hadrons/Lattice QCD. We thank the Swiss Center for
Scientific Computing in Manno, the University of Regensburg and LRZ M\"unchen for
computational resources.

\end{document}